\documentclass[journal]{IEEEtran/IEEEtran}
\usepackage[latin1]{inputenc}
\usepackage[T1]{fontenc}
\usepackage[english]{babel}
\usepackage{multirow}
\usepackage{soul}

\usepackage{cite} \makeatletter \let\MYcaption\@makecaption
\makeatother \usepackage{caption}
\usepackage[font=footnotesize]{subcaption} \makeatletter
\let\@makecaption\MYcaption

\makeatother \usepackage{balance}

\usepackage[dvips]{graphicx}
\graphicspath{{img_arxiv/}}
 \DeclareGraphicsExtensions{.eps}
\usepackage[dvipsnames]{xcolor}

\usepackage{amsmath}
\usepackage{amssymb}
\usepackage{amsfonts}

\usepackage{url}
\usepackage{hyperref}
\hypersetup{
    unicode=false,          
    pdftoolbar=true,        
    pdfmenubar=true,        
    pdffitwindow=false,     
    pdfstartview={FitH},    
    pdftitle={My title},    
    pdfauthor={Author},     
    pdfsubject={Subject},   
    pdfcreator={Creator},   
    pdfproducer={Producer}, 
    pdfkeywords={keyword1, key2, key3}, 
    pdfnewwindow=true,      
    colorlinks=true,       
    linkcolor=black,          
    urlbordercolor={1 1 1},
    citecolor=black,        
    filecolor=magenta,      
    urlcolor=black           
}

\usepackage[binary-units=true]{siunitx}
\DeclareSIUnit{\sample}{S}
\DeclareSIUnit{\baud}{Bd}

\newcommand{\orcid}[1]{\href{https://orcid.org/#1}{\includegraphics[width=10px]{ORCIDiD.eps}}}
\begin{document}

\title{On the Application of Error Backpropagation to the Background
  Calibration of Time Interleaved ADC for Digital Communication
  Receivers} \author{\IEEEauthorblockN{ Fredy
    Solis\IEEEauthorrefmark{2}, Benjam\'in
    T. Reyes\IEEEauthorrefmark{2}, Dami\'an A.
    Morero\IEEEauthorrefmark{3}, and
    Mario R. Hueda\IEEEauthorrefmark{3}} \\
  \IEEEauthorblockA {
    \IEEEauthorrefmark{2} Fundaci\'on Fulgor - Romagosa 518 - C\'ordoba (5000) - Argentina \\
    \IEEEauthorrefmark{3} Laboratorio de Comunicaciones Digitales - Universidad Nacional de C\'ordoba - CONICET\\
    Av. V\'elez Sarsfield 1611 - C\'ordoba (X5016GCA) - Argentina\\
    Email: \{benjamin.reyes, mario.hueda\}@unc.edu.ar}}


\maketitle

\begin{abstract}
  This paper introduces a backpropagation-based technique for the
  calibration of the mismatch errors of time-interleaved analog to
  digital converters (TI-ADCs). This technique is applicable to
  digital receivers such as those used in coherent optical
  communications. The error at the slicer of the receiver is processed
  using a modified version of the well known backpropagation algorithm
  from machine learning. The processed slicer error can be directly
  applied to compensate the TI-ADC mismatch errors with an adaptive
  equalizer, or it can be used to digitally estimate and correct said
  mismatch errors using analog techniques such as delay cells and
  programmable gain amplifiers (PGA). The main advantages of the
  technique proposed here compared to prior art are its robustness,
  its speed of convergence, and the fact that it always works in
  background mode, independently of the oversampling factor and the
  properties of the input signal, as long as the receiver
  converges. Moreover, this technique enables the joint compensation
  of impairments not addressed by traditional TI-ADC calibration
  techniques, such as I/Q skew in quadrature modulation
  receivers. Simulations are presented to demonstrate the
  effectiveness of the technique, and low complexity implementation
  options are discussed.
\end{abstract}


\IEEEpeerreviewmaketitle
\section{Introduction}
\label{sec:intro}

\IEEEPARstart{H}{igh} speed digital receivers such as those used in
coherent optical
communications~\cite{morero_design_2016,faruk_digital_2017,
  fludger_digital_2014,
  crivelli_adaptive_2004,crivelli_architecture_2014,
  agrell_roadmap_2016, roberts_high_2015,kikuchi_fundamentals_2016,
  bosco_advanced_2019} require high bandwidth, high sampling rate
analog to digital converters (ADC).  Current coherent receivers
operate at symbol rates around 96 Giga-baud (GBd) and require ADC
bandwidths of about \SI{50}{\giga Hz} and sampling rates close to
\SI{150}{\GHz}.  In the near future symbol rates will increase to
\SIrange{128}{150}{\giga\baud} or higher, required bandwidths will be
in the range of \SIrange{65}{75}{\GHz}, and sampling rates in the
\SIrange{200}{250}{\GHz} range.  The technique universally applied so
far to achieve these high bandwidths and sampling rates in coherent
transceivers is the time interleaved ADC (TI-ADC)
\cite{laperle_advances_2014,kull_cmos_2016}. Frequency interleaved
ADCs (FI-ADC) may become a promising alternative in the near future
\cite{Passetti20-ISCAS}.

The performance of TI-ADCs is affected by mismatches among the
interleaves, particularly the mismatches of sampling time, gain, and
DC offset\cite{kurosawa_explicit_2001,vogel_impact_2005}.  There has
been a large body of literature dedicated to various techniques to
calibrate these mismatches~\cite{reyes_joint_2012,lin_10b_2016,
  reyes_design_2017,el-chammas_12-gs/s_2011,wei_8_2014,song_10-b_2017,
  haftbaradaran_background_2008, elbornsson_blind_2005,
  salib_low-complexity_2017,
  matsuno_all-digital_2013,devarajan_12-b_2017,
  lee_1_2014,mafi_digital_2017, ali_14-bit_2016,
  harpe_oversampled_2014,luna_compensation_2006}.  For a thorough
review and discussion of previous TI-ADC calibration techniques,
please see \cite{murmann_digitally_2013,harpe_digitally_2015} and
references therein.  In general, existing techniques suffer from one
or more of the following drawbacks: \textit{i)} Dependence on the
properties of the input signal or the oversampling factor for proper
operation\cite{wei_8_2014,song_10-b_2017,elbornsson_blind_2005,salib_low-complexity_2017,haftbaradaran_background_2008,song_10-b_2018};
\textit{ii)} Requiring an extra ADC to provide a
reference\cite{lin_10b_2016,el-chammas_12-gs/s_2011}; \textit{iii)}
Introducing intentional degradations (e.g., \textit{dither}) in the
ADC output in order to find the calibration
parameters\cite{matsuno_all-digital_2013,devarajan_12-b_2017}, and
\textit{iv)} Slow
convergence\cite{reyes_joint_2012,reyes_design_2017}.  In this paper
we propose a new background technique that overcomes the above
limitations, and is applicable to receivers for digital communications
such as those used for coherent optical transmission
\cite{morero_design_2016}. The basic idea consists in the use of a low
complexity adaptive equalizer, called \textit{Compensation Equalizer}
(CE), to compensate the mismatches of the TI-ADC. The CE is adapted
using the stochastic gradient descent (SGD) algorithm and a post
processed version of the error available at the slicer of the
receiver, where the post processing is done using the backpropagation
algorithm \cite{rumelhart_learning_1986,goodfellow_deep_2016}.

Next we discuss some of the state-of-the-art TI-ADC calibration
techniques and compare them to the one proposed in this paper.  The
approaches developed in~\cite{wei_8_2014, salib_low-complexity_2017,
  lin_10b_2016, haftbaradaran_background_2008, elbornsson_blind_2005}
use the autocorrelation of the quantized signal to estimate the timing
mismatches and adjust them in the analog or digital domain.  One
serious limitation of this technique is related to the properties of
the input signal~\cite{wei_8_2014, haftbaradaran_background_2008},
which cause the calibration algorithm to diverge for some particular
input frequencies (given by $f_{in}=m\frac{f_s}{2M}$, with
$m\in \left\{0,\cdots,M-1\right\}$ where $f_s$ is the sampling rate
and $M$ is the number of interleaves of the TI-ADC). This makes this
technique not suitable for receivers with a particular oversampling
ratio (OSR).

Other techniques perform the calibration based on statistical
properties of the quantized
signal~\cite{song_10-b_2017,lee_1_2014,mafi_digital_2017,song_10-b_2018}.
Histograms of sub-ADC outputs are created and calculations such as the
variance or cumulative distribution function are made to estimate the
error introduced by the timing mismatch. The effectiveness of these
approaches depends on the input signal characteristics and therefore
their robustness is problematic.  In other examples,
\cite{lee_1_2014,song_10-b_2017} an auxiliary channel (working at the
overall sampling rate of the TI-ADC) is required to provide a
reference or to enable the estimation, respectively.

Dither injection techniques are based on the addition of a signal in
either the analog or digital domain to facilitate the estimation of
the calibration parameters~\cite{devarajan_12-b_2017, ali_14-bit_2016,
  morie_71db-sndr_2013, harpe_oversampled_2014}.  They are often used
in high resolution ADCs ($>10$ \emph{effective number of bits} or
ENOB).  Depending on the level of the dither signal, this technique
can limit the swing of the input signal.  One of the most common
limitations of existing techniques is their inability to adjust
calibration parameters of \textit{different} nature simultaneously. A
technique where calibration of a given impairment does not depend on
calibration of the other impairments is highly desirable. For example,
the stability of the calibration of the timing mismatch should not
rely on how free from offset, gain or bandwidth mismatches the sampled
signal is.

This work presents a new technique based on adaptive equalization
\cite{solis_background_2020} that runs inherently in background and is
able to overcome the aforementioned limitations.  The proposed
technique is intended to be applied in high-speed receivers for
digital communication systems such as those used in coherent optical
transmission \cite{morero_design_2016}.

Adaptive equalization has been shown \cite{luna_compensation_2006,
  agazzi_90_2008} to be a powerful technique to compensate errors in
TI-ADCs. However, its application to some types of receivers for
digital communications, particularly coherent optical transceivers,
has been limited by the lack of availability of a suitable error
signal to use in the SGD algorithm. In \cite{luna_compensation_2006,
  agazzi_90_2008} the equalizer used to compensate TI-ADC impairments
is the main receiver equalizer, or \textit{Feedforward Equalizer}
(FFE). This is possible in the referenced works because the FFE is
immediately located after the TI-ADC, without any other blocks in
between. Therefore the FFE can access and compensate directly the
impairments of the different interleaves. Also, the slicer error
carries information about the impairments of the individual
interleaves and therefore the FFE adaptation algorithm can drive its
coefficients to a solution that jointly compensates the channel and
the TI-ADC impairments. In the case of coherent optical receivers
there is at least one block, the \textit{Bulk Chromatic Dispersion
  Equalizer} (BCD) between the TI-ADC and the FFE. The BCD causes
signal components associated with different interleaves of the TI-ADC
to be mixed in a way that makes the use of the FFE unsuitable to
compensate them. Therefore a \textit{separate} equalizer, immediately
located after the TI-ADC, is necessary. This is the previously
mentioned CE. Although in this way the CE has direct access to the
impairments of the different TI-ADC interleaves, the slicer error is
not directly applicable to adapt it, because error components
associated with different TI-ADC have also been mixed by the BCD (and
possibly other signal processing blocks, depending on the architecture
of the receiver).  This paper solves that problem through the use of
the backpropagation algorithm ~\cite{rumelhart_learning_1986}, an
algorithm widely used in machine
learning~\cite{goodfellow_deep_2016}. Its main characteristic is that,
in a multi-stage processing chain where several cascaded blocks have
adaptive parameters, it is able to determine the error generated by
each one of these sets of parameters for all the
stages. Backpropagation is used in combination with the SGD algorithm
to adjust the parameters of the CE in order to minimize the Mean
Squared Error (MSE) at the slicer of the receiver.  The use of the CE
in combination with the backpropagation algorithm results in robust,
fast converging background compensation or calibration. As mentioned
before, the compensation is not limited to the impairments of
individual TI-ADCs (which is the case for traditional calibration
techniques), but it extends itself to the entire receiver analog front
end, enabling the compensation of impairments such as time skew
between the in-phase and the quadrature components of the signal in a
receiver based on Quadrature Amplitude Modulation (QAM) or Phase
Modulation (PM).

In the architecture just described, the compensation is achieved by an
all-digital technique. A variant of the technique based on a
mixed-signal calibration is also proposed in this paper. In this
variant, the backpropagation and the SGD algorithms are used to
estimate the TI-ADC mismatch errors, but the equalizer per se is not
built.

Because ultrafast adaptation is usually not necessary, the
backpropagation algorithm can be implemented in a highly subsampled
hardware block which does not require parallel processing. Therefore
the implementation complexity of the proposed technique is low, as
discussed in detail in Section \ref{s:complexity}.

The rest of this paper is organized as follows. Section
\ref{sec:System Model} presents a discrete time model of the TI-ADC
system in a dual polarization optical coherent receiver. The error
backpropagation based adaptive compensation equalizer is introduced in
Section \ref{sec:EBP}. Simulation results are presented and discussed
in Section \ref{s:sim_res}. The hardware complexity of the proposed
compensation scheme is discussed in Section \ref{s:complexity}, and
conclusions are drawn in Section \ref{s:conclusion}.

\section{System Model}
\label{sec:System Model}

\begin{figure}
  \centering
  \includegraphics[width=\columnwidth]{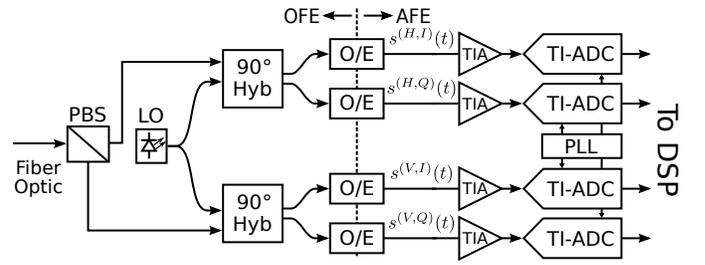}
  \caption{\label{f:ofe} Optical/analog front-end for a TI-ADC-based
    coherent optical receiver. The optical signal is split into four
    electrical lanes that are converted by a TI-ADC. \emph{PBS}:
    polarization beam splitter; \emph{LO}: local oscillator;
    \emph{$90^o$ Hyb}: $90^o$ hybrid coupler.}
\end{figure}

Although the new compensation algorithm proposed in this work can be
applied to any high-speed digital communication receiver, to make the
discussion more concrete we focus the study on dual-polarization (DP)
optical coherent transceivers
\cite{morero_design_2016,faruk_digital_2017, fludger_digital_2014,
  crivelli_adaptive_2004,crivelli_architecture_2014}. A block diagram
of an optical front-end (OFE) for a DP coherent receiver is shown in
Fig.~\ref{f:ofe}. The optical input signal is decomposed by the OFE to
obtain four components, the in-phase and quadrature (I/Q) components
of the two polarizations (H/V).  The photodetectors convert the
optical signals to photocurrents which are amplified by
trans-impedance amplifiers (TIAs). The analog front-end (AFE) is in
charge of the acquisition and conversion of the electrical signal to
the digital domain. Typically, oversampled digital receivers are used
to compensate the dispersion experienced in optical links (e.g.,
$T_s=\frac{T}{2}$ where $T_s$ and $T$ are the sampling and symbol
periods, respectively)~\cite{crivelli_adaptive_2004}. Next we develop
the model of the optical channel used in the remainder of this paper.

Let
$a_k^{({\mathcal P})}=a_k^{({\mathcal P},{ I})}+ja_k^{({\mathcal
    P},{Q})}$ be the $k$-th quadrature amplitude modulation (QAM)
symbol in polarization ${\mathcal P}\in\{H,V\}$. An optical fiber link
with chromatic dispersion (CD) and polarization-mode dispersion (PMD)
can be modeled as a $2 \times 2$ \emph{multiple-input multiple output}
(MIMO) complex-valued channel \cite{crivelli_adaptive_2004}
encompassing four complex filters with impulse responses
${\overline h}_{m,n}(t)$ where $m,n=1,2$. Then, the received
noise-free electrical signals provided by the optical demodulator can
be expressed as \cite{crivelli_adaptive_2004}
\begin{align}
  \label{eq:eq0H}
  s^{(H)}(t)&=s^{(H,I)}(t)+js^{(H,Q)}(t)\\
  \nonumber
            &=e^{j\omega_0 t}\left[\sum_k a_k^{(H)}{\overline h}_{1,1}(t-kT)+ a_k^{(V)}{\overline h}_{1,2}(t-kT)\right],\\
  \label{eq:eq0V}
  s^{(V)}(t)&=s^{(V,I)}(t)+js^{(V,Q)}(t)\\
  \nonumber
            &=e^{j\omega_0t}\left[\sum_k a_k^{(H)}{\overline h}_{2,1}(t-kT)+ a_k^{(V)}{\overline h}_{2,2}(t-kT)\right],
\end{align}
where $1/T$ is the symbol rate and $\omega_0$ is the optical carrier
frequency offset.

\subsection{Discrete-Time Model of the AFE and the TI-ADC}

\begin{figure}
  \centering
  \includegraphics[width=\columnwidth]{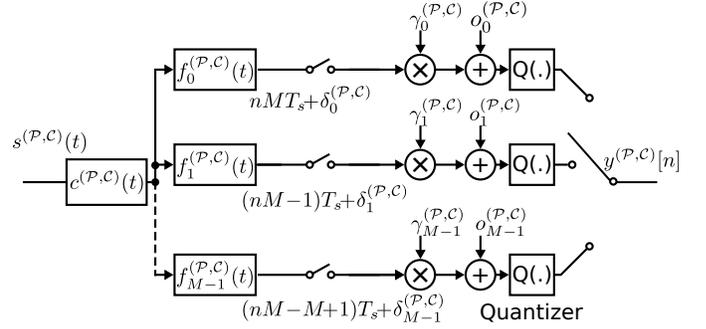}
  \caption{\label{f:fig1} Analog front-end for polarization
    $\mathcal P \in\{H,V\}$ and component $\mathcal C \in\{I,Q\}$ in a
    TI-ADC-based DP coherent optical receiver.}
\end{figure}

In this section we introduce a discrete-time model for the AFE and
TI-ADC system of Fig. \ref{f:ofe} including their impairments. A
simplified model of the analog path for one component
${\mathcal C}\in\{I,Q\}$ in a given polarization
${\mathcal P\in\{H,V\}}$ is shown in Fig.~\ref{f:fig1}.  Each lane of
the AFE includes a filter with impulse response
$c^{({\mathcal P},{\mathcal C})}(t)$ that models the response of the
electrical interconnections between the optical demodulator and the
TIA, the TIA response itself, and any other components in the signal
path up to an $M$-parallel TI-ADC system. Mismatches between
$c^{({\mathcal P},I)}(t)$ and $c^{({\mathcal P},Q)}(t)$ may cause time
delay or \emph{skew} between components $I$ and $Q$ of a given
polarization $\mathcal P$, which degrade the receiver performance. As
we shall show here, the proposed background calibration algorithm is
able to compensate not only the imperfections of the TI-ADC, but also
the I/Q skew and any other mismatches among the signal paths.

The independent frequency responses of the $M$ track and hold units in
an $M$-channel TI-ADC system are modeled by blocks
$f^{({\mathcal P},{\mathcal C})}_{m}(t)$ with $m=0,\cdots, M-1$. Each
$M$-way interleaved TI-ADC path is sampled every $M/f_s=MT_s$ seconds
with a proper sampling phase. Parameters
$\delta_m^{({\mathcal P},{\mathcal C})}$ and
$o^{({\mathcal P},{\mathcal C})}_m$ model the sampling time errors and
the DC offsets, respectively. Path gains are modeled by
\begin{equation}
  \label{eq:gain_error}
  \gamma^{({\mathcal P},{\mathcal C})}_m=1+\Delta_{\gamma^{({\mathcal
        P},{\mathcal C})}_m},
\end{equation}
where $\Delta_{\gamma^{({\mathcal P},{\mathcal C})}_m}$ is the gain
error.

\begin{figure}
  \centering
  \includegraphics[width=\columnwidth]{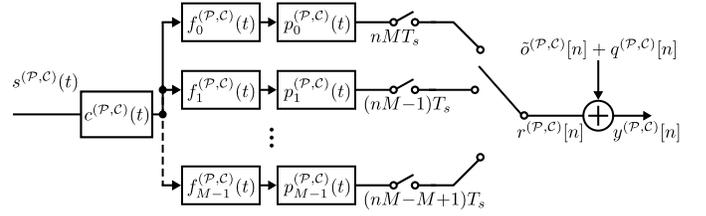}
  \caption{\label{f:fig2} Modified model of the analog front-end and
    TI-ADC for polarization $\mathcal P \in\{H,V\}$ and component
    $\mathcal C \in\{I,Q\}$ in a DP coherent optical receiver.}
\end{figure}

Following \cite{luna_compensation_2006,agazzi_90_2008}, the sampling
phase error $\delta_m^{({\mathcal P},{\mathcal C})}$ and the path gain
$\gamma_m^{({\mathcal P},{\mathcal C})}$ are modeled by an analog
interpolation filter with impulse response
$p_m^{({\mathcal P},{\mathcal C})}(t)$ followed by ideal sampling as
depicted in Fig. \ref{f:fig2}. Assuming that the bit-resolution of the
ADC's is sufficiently high, the quantizer can be modeled as additive
white noise with uniform distribution. Also, at high-frequency (i.e.,
$1/T_s$), the offsets $o^{({\mathcal P},{\mathcal C})}_m$ generate an
$M$-periodic signal denoted as
${\tilde o}^{({\mathcal P},{\mathcal C})}[n]$ such that
${\tilde o}^{({\mathcal P},{\mathcal C})}[n]={\tilde o}^{({\mathcal
    P},{\mathcal C})}[n+M]$ with
\begin{equation}
  \label{eq:tilde)}
  {\tilde o}^{({\mathcal P},{\mathcal C})}[m]={o}^{({\mathcal
      P},{\mathcal C})}_m,\quad m=0,\cdots,M-1.
\end{equation}

Then, the digitized high-frequency samples can be expressed as
\begin{equation}
\label{eq:eq1}
y^{({\mathcal P},{\mathcal C})}[n]=r^{({\mathcal P},{\mathcal C})}[n]+{\tilde o}^{({\mathcal P},{\mathcal C})}[n]+q^{({\mathcal P},{\mathcal C})}[n],
\end{equation}
where $r^{({\mathcal P},{\mathcal C})}[n]$ is the signal component
provided by the $M$-channel TI-ADC, and
$q^{({\mathcal P},{\mathcal C})}[n]$ is the quantization noise (see
Fig. \ref{f:fig2}).

\begin{figure}
  \centering
  \includegraphics[width=\columnwidth]{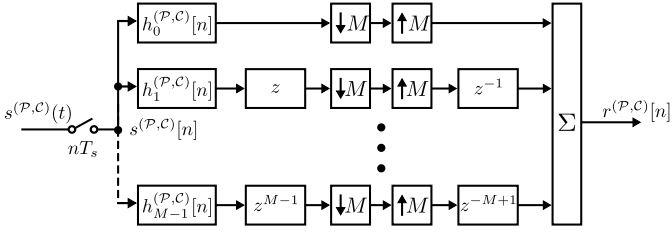}
  \caption{\label{f:fig3} Equivalent discrete-time model of the analog
    front-end and TI-ADC system with impairments for the signal
    component given by \eqref{eq:eq4} (i.e., without DC offsets and
    quantization noise) for polarization $\mathcal P \in\{H,V\}$ and
    component $\mathcal C \in\{I,Q\}$.}
\end{figure}

We define the total impulse response of a given subchannel as
\begin{equation}
\label{eq:eq2}
h_m^{({\mathcal P},{\mathcal C})}(t)=c^{({\mathcal P},{\mathcal C})}(t)\otimes f_m^{({\mathcal P},{\mathcal C})}(t) \otimes p_m^{({\mathcal P},{\mathcal C})}(t),
\end{equation}
where $m=0,\cdots, M-1$ and $\otimes$ denotes the convolution
operation. Let $H_m^{({\mathcal P},{\mathcal C})}(j\omega)$ and
$S^{({\mathcal P},{\mathcal C})}(j\omega)$ be the Fourier transforms
(FTs) of $h_m^{({\mathcal P},{\mathcal C})}(t)$ and
$s^{({\mathcal P},{\mathcal C})}(t)$, respectively. In digital
communication systems with spectral shaping is
$|S^{({\mathcal P},{\mathcal C})}(j\omega)|\approx 0$ for
$|\omega|\ge \pi/T_s$. Further assuming that
$|H_m^{({\mathcal P},{\mathcal C})}(j\omega)|\approx 0$ for
$|\omega|\ge \pi/T_s$, the analog filtering of Fig. \ref{f:fig2} can
be represented by a real discrete-time model as depicted in
Fig. \ref{f:fig3} by using
\begin{equation}
\label{eq:eq3}
h^{({\mathcal P},{\mathcal C})}_m[n]=T_sh_m^{({\mathcal P},{\mathcal C})}(nT_s),\quad m=0,\cdots,M-1.
\end{equation}
Therefore it can be shown that the digitized high-frequency signal can
be expressed as:
\begin{equation}
\label{eq:eq4}
r^{({\mathcal P},{\mathcal C})}[n]=\sum_{l} {\tilde h}^{({\mathcal P},{\mathcal C})}_n[l] s^{({\mathcal P},{\mathcal C})}[n-l],
\end{equation}
where
$s^{({\mathcal P},{\mathcal C})}[n]=s^{({\mathcal P},{\mathcal
    C})}(nT_s)$ and ${\tilde h}^{({\mathcal P},{\mathcal C})}_n[l]$ is
the impulse response of a time-varying filter, which is an
$M$-periodic sequence such
${\tilde h}^{({\mathcal P},{\mathcal C})}_n[l]={\tilde h}^{({\mathcal
    P},{\mathcal C})}_{n+M}[l]$, and defined by
\begin{equation}
  \label{eq:eq5}
  {\tilde h}^{({\mathcal P},{\mathcal C})}_n[l]={h}^{({\mathcal P},{\mathcal C})}_n[l], \quad n=0,\cdots,M-1,
  \forall l,
\end{equation}
with ${h}^{({\mathcal P},{\mathcal C})}_n[l]$ given by
\eqref{eq:eq3}\footnote{See \cite{saleem_adaptive_2010} and references
  therein for more details about this formulation.}. We highlight that
\eqref{eq:eq4} includes the impact of both the AFE mismatches and the
$M$-channel TI-ADC impairments. Replacing \eqref{eq:eq4} in
\eqref{eq:eq1}, the digitized high-frequency sequences result
\begin{align}
  \nonumber
  y^{({\mathcal P},{\mathcal C})}[n]=&\sum_{l} {\tilde h}^{({\mathcal P},{\mathcal C})}_n[l] s^{({\mathcal P},{\mathcal C})}[n-l]+{\tilde o}^{({\mathcal P},{\mathcal C})}[n]+\\
  \label{eq:eq1b}
              &q^{({\mathcal P},{\mathcal C})}[n].
\end{align}

\subsection{Compensation of AFE Mismatch and TI-ADC Impairments}
Similar to what was done in previous works
\cite{luna_compensation_2006,agazzi_90_2008, saleem_adaptive_2010}, we
propose to use an adaptive digital compensation filter applied after
the mitigation of the offset sequence, i.e.,
\begin{equation}
  \label{eq:eq6}
  x^{({\mathcal P},{\mathcal C})}[n]=\sum_{l=0}^{L_g-1} {\tilde g}^{({\mathcal P},{\mathcal C})}_n[l]
  {w}^{({\mathcal P},{\mathcal C})}[n-l],
\end{equation}
where ${\tilde g}^{({\mathcal P},{\mathcal C})}_n[l]$ is the
$M$-periodic time-varying impulse response of the compensation filter
(i.e.,
${\tilde g}^{({\mathcal P},{\mathcal C})}_n[l]={\tilde g}^{({\mathcal
    P},{\mathcal C})}_{n+M}[l]$), $L_g$ is the number of taps of the
compensation filters, and $w^{({\mathcal P},{\mathcal C})}[n]$ is the
offset compensated signal given by
\begin{equation}
  \label{eq:w}
  w^{({\mathcal P},{\mathcal C})}[n]=y^{({\mathcal P},{\mathcal
      C})}[n]-{\hat {\tilde o}}^{({\mathcal P},{\mathcal C})}[n],
\end{equation}
with ${\hat {\tilde o}}^{({\mathcal P},{\mathcal C})}[n]$ being the
estimated $M$-periodic offset sequence. The combination of the offset
compensation blocks and the compensation filters
${\tilde g}^{({\mathcal P},{\mathcal C})}_n[l]$ constitutes the
\emph{Compensation Equalizer} (Fig. \ref{f:f5}).

A proper strategy to estimate the response of the CE is
required. Notice that adaptive calibration techniques based on a
reference ADC such as in \cite{saleem_adaptive_2010} cannot be used to
compensate mismatches between the $I$ and $Q$ signal paths. In the
following we propose the \emph{backpropagation} technique to adapt the
CE.

\section{Error Backpropagation Based Compensation of AFE and TI-ADC
  Impairments in DP Optical Coherent Receivers}
\label{sec:EBP}
\begin{figure}[t]
  \centering
  \includegraphics[width=\columnwidth]{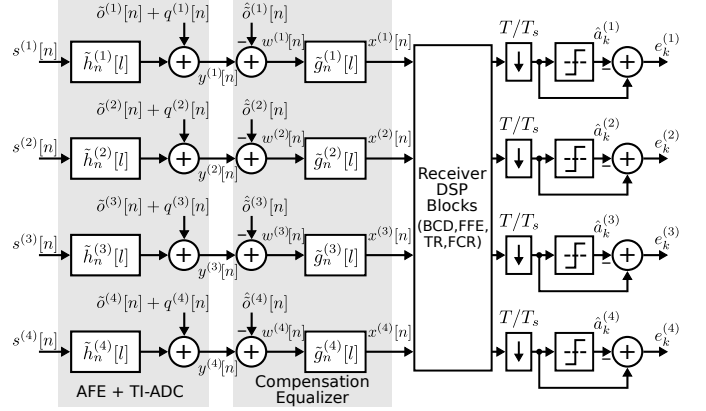}
  \caption{\label{f:f5} Block diagram of a dual-polarization optical
    coherent receiver with the \emph{Compensation Equalizer} (CE) for
    mitigating the effects of both AFE mismatches and TI-ADC
    impairments.}
\end{figure}

Based on the previous analysis, Fig. \ref{f:f5} depicts a block
diagram of the AFE+TI-ADC in a dual-polarization optical coherent
receiver with the adaptive calibration block, which includes four
instances of the real filter as defined by \eqref{eq:eq6}. For
simplicity, we modified the notation of the system model of
Fig. \ref{f:fig3}. Note that we use an integer index between 1 and 4
to represent a certain component in a given polarization:
$``(1) "=(H,I), ``(2)"=(H,Q), ``(3)"=(V,I)$, and $``(4)"=(V,Q)$.

The main receiver functions are included in the digital signal
processing (DSP) block of Fig. \ref{f:f5}, which works with samples
every $T_s$ seconds. In summary, some of the most important DSP
algorithms used in these receivers are the chromatic dispersion
equalizer (or BCD), the MIMO FFE to compensate the polarization-mode
dispersion, \emph{Timing Recovery} (TR) from the received symbols, the
\emph{Fine Carrier Recovery} (FCR) to compensate the carrier phase and
frequency offset, and the \emph{Forward Error Correction} (FEC)
decoder. Readers interested in more details on optical coherent
receivers can see \cite{morero_design_2016, faruk_digital_2017,
  fludger_digital_2014} and references therein.

\subsection{All Digital Compensation Architecture}
Let ${g}^{(i)}_m[l]$ with $i=1,\cdots,4$ be the filter impulse
response ${\tilde g}^{(i)}_{n}[l]$ in one period defined as
\begin{equation}
  \label{eq:gPC}
  {g}^{(i)}_m[l]={\tilde g}^{(i)}_{m+n_0}[l], \quad m=0,\cdots,M-1,
\end{equation}
where $l=0,\cdots,L_g-1$ and $n_0$ is an arbitrary time index multiple
of $M$. The filter taps of the CE ${g}^{(i)}_m[l]$ are adapted by
using the slicer error at the output of the receiver DSP block. Let
$e_k^{(j)}$ be the \emph{slicer error} defined by
\begin{equation}
  \label{eq:ePC}
  e_k^{(j)}=u_k^{(j)}-{\hat a}_k^{(j)},\quad j=1,\cdots,4,
\end{equation}
where $u_k^{(j)}$ is the input of the slicer and ${\hat a}_k^{(j)}$ is
the $k$-th detected symbol at the slicer output (see
Fig. \ref{f:f5}). Notice that the sampling rate of the slicer inputs
$u_k^{(j)}$ is $1/T$, therefore a subsampling of $T/T_s$ is carried
out after the receiver DSP block. Then, we define the total squared
error at the slicer as
\begin{equation}
  \label{eq:eT}
  {\mathcal E}_k=\sum_{j=1}^4|e_k^{(j)}|^2.
\end{equation}

Let $E\{{\mathcal E}_k\}$ be the MSE at the slicer with $E\{.\}$
denoting the expectation operator. In this work we use the \emph{least
  mean squares} (LMS) algorithm to iteratively adapt the real
coefficients of the CE given by \eqref{eq:gPC}, in order to minimize
the MSE at the slicer:
\begin{equation}
  \label{eq:lmsg}
  {\bold g}^{(i)}_{m,p+1}={\bold g}^{(i)}_{m,p}- \beta
  \nabla_{{\bold g}^{(i)}_{m,p}} E\{{\mathcal
    E}_k\},
\end{equation}
where $i=1,\cdots,4,$; $m=0,\cdots,M-1$; $p$ denotes the number of
iteration, ${\bold g}^{(i)}_{m,p}$ is the $L_g$-dimensional
coefficient vector at the $p$-th iteration given by
\begin{equation}
  \label{eq:vgPC}
  {\bold g}^{(i)}_{m,p}=\left[{g}^{(i)}_{m,p}[0],{g}^{(i)}_{m,p}[1],\cdots,{g}^{(i)}_{m,p}[L_g-1]  \right]^T;
\end{equation}
$\beta$ is the adaptation step, and
$\nabla_{{\bold g}^{(i)}_{m,p}} E\{{\mathcal E}_k\}$ is the gradient
of the MSE with respect to the filter vector ${\bold g}^{(i)}_{m,p}$.

We emphasize that the computation of the MSE gradient is not trivial
since ${\mathcal E}_k$ is not the error at the output of the CE
block. To get the proper error samples to adapt the coefficients of
the filters as expressed in \eqref{eq:lmsg}, we propose the
\emph{backpropagation algorithm} widely used in \emph{machine
  learning} \cite{rumelhart_learning_1986,
  goodfellow_deep_2016}. Towards this end, the slicer errors are
backpropagated as described in~\cite{morero_forward_2018}. Finally,
based on these backpropagated errors we can estimate the gradient
$\nabla_{{\bold g}^{(i)}_{m,p}} E\{{\mathcal E}_k\}$ as usual in the
classical LMS algorithm.

\subsection{Error Backpropagation (EBP)}
\label{s:bp_formulation}
\begin{figure}
  \centering \includegraphics[width=\columnwidth]{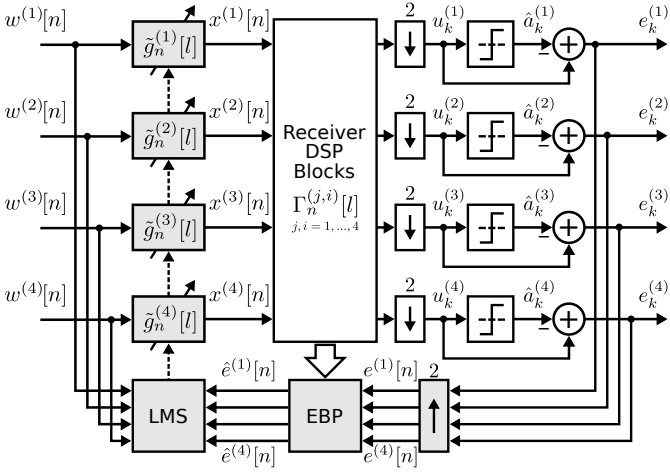}
  \caption{\label{f:f6} Block diagram of the proposed \emph{error
      backpropagation} (EBP) based adaptation architecture for
    AFE+TI-ADC impairment compensation in a dual-polarization optical
    coherent receiver with $T/T_s=2$.}
\end{figure}

Without loss of generality, we assume that the receiver DSP block can
be modeled as a real time-varying $4 \times 4$ MIMO $T/2$ fractional
spaced equalizer (i.e., $T_s=T/2$), which is able to compensate CD and
PMD among other optical fiber channel effects. Then, the downsampled
output of the $T/2$ receiver DSP block can be written as (see Fig
\ref{f:f6})
\begin{equation}
  \label{eq:u1}
  u^{(j)}_k=\sum_{i=1}^4\sum_{l=0}^{L_{\Gamma}-1}
  {\Gamma}^{(j,i)}_{2k}[l]{x}^{(i)}[2k-l],\quad j=1,\cdots,4,
\end{equation}
where ${\Gamma}_n^{(j,i)}[l]$ is the time-varying impulse response of
the filter with input $i$ and output $j$, $L_{\Gamma}$ is the number
of taps of the filter, while ${x}^{(i)}[l]$ is the signal at the DSP
block input $i$ given by \eqref{eq:eq6}, i.e.,
\begin{equation}
  \label{eq:eq6b}
  x^{(i)}[n]=\sum_{l'=0}^{L_g-1} {g}^{(i)}_{\lfloor n\rfloor_M}[l']
  {w}^{(i)}[n-l'],\quad i=1,\cdots,4,
\end{equation}
where ${g}^{(i)}_m$ is the impulse response defined by \eqref{eq:gPC},
$\lfloor .\rfloor_M$ denotes the modulo $M$ operation, and
${w}^{(i)}[n]$ is the DC compensated signal \eqref{eq:w}.

As usual with the SGD based adaptation, we replace the gradient of the
MSE, $\nabla_{{\bold g}^{(i)}_{m,p}} E\{{\mathcal E}_k\}$, by a noisy
estimate, $\nabla_{{\bold g}^{(i)}_{m}} {\mathcal E}_k$. In the
Appendix we show that an \emph{instantaneous} gradient of the squared
error \eqref{eq:eT} can be expressed as
\begin{equation}
  \label{eq:grad}
  \nabla_{{\bold g}^{(i)}_{m}} {\mathcal E}_k=\alpha {\hat e}^{(i)}[m+kM]{\bold w}^{(i)}[m+kM],
\end{equation}
where $\alpha$ is a certain constant, ${\bold w}[n]$ is the
$L_g$-dimensional vector with the samples at the CE input, i.e.,
\begin{equation}
  \label{eq:vecw}
  {\bold w}^{(i)}[n]=\left[{w}^{(i)}[n],{w}^{(i)}[n-1],\cdots,{w}^{(i)}[n-L_g+1]  \right]^T,
\end{equation}
while ${\hat e}^{(i)}[n]$ is the \emph{backpropagated error} given
by
\begin{equation}
  \label{eq:bpe}
  {\hat e}^{(i)}[n]=\sum_{j=1}^4\sum_{l=0}^{L_{\Gamma}-1}\Gamma^{(j,i)}_{n+l}[l] e^{(j)}[n+l],
\end{equation}
with $e^{(j)}[n]$ being the \emph{oversampled} slicer error obtained
from the \emph{baud-rate} slicer error $e_k^{(j)}$ in \eqref{eq:ePC}
as
\begin{equation}
  \label{eq:oe}
  e^{(j)}[n] = 
  \begin{cases} 
    e_{n/2}^{(j)}              & \mbox{if } n= 0,\pm 2,\pm 4,\cdots   \\
    0 & \mbox{otherwise}
  \end{cases}.
\end{equation}
Then, a full digital compensation architecture can be derived by using
an adaptive CE with
\begin{equation}
  \label{eq:lmsg2}
  {\bold g}^{(i)}_{m,p+1}={\bold g}^{(i)}_{m,p}- \mu \nabla_{{\bold g}^{(i)}_{m,p}} {\mathcal E}_k,
\end{equation}
where $\mu=\alpha \beta$ is the step-size. Furthermore, based on the
backpropagated error \eqref{eq:bpe} it is possible to estimate the DC
offsets in the input samples as follows
\begin{equation}
  {\hat {o}}^{(i)}_{p+1}[m]={\hat {o}}^{(i)}_{p}[m]-\mu_o{\hat
    e}^{(i)}[n+m], \quad m=0,\cdots, M-1,
  \label{eq:tildeo}
\end{equation}
where ${\hat {o}}^{(i)}_{p}[m]$ is the estimate at the $p$-th
iteration of the DC offset sequence in one period (see \eqref{eq:w}),
and $\mu_o$ is the step-size of the DC offset estimator.

Competition between the CE and any adaptive DSP blocks in
$\Gamma_n^{j,i}[l]$ (e.g., the FFE) may generate instability,
therefore an adaptation constraint must be included. For example, one
of the $4M$ sets of the filter coefficients can be limited to only be
a time delay line, for example, ${g}^{(0)}_{0}[l]=\delta_{l,l_d}$
where $l=0,\cdots,L_g-1$ and $l_d=\frac{L_g+1}{2}$ ($L_g$ is assumed
odd).

Since channel impairments change slowly over time, the coefficient
updates given by \eqref{eq:lmsg2} and \eqref{eq:tildeo} do not need to
operate at full rate, and subsampling can be applied. The latter
allows implementation complexity to be significantly
reduced. Additional complexity reduction is enabled by: 1) strobing
the algorithms once they have converged, and/or 2) implementing them
in firmware in an embedded processor, typically available in coherent
optical transceivers. Practical aspects of the hardware implementation
shall be discussed in Section \ref{s:complexity}.

\subsection{Mixed-Signal Compensation Architecture}
\label{sec:mixed}
\begin{figure}
  \centering
  \includegraphics[width=\columnwidth]{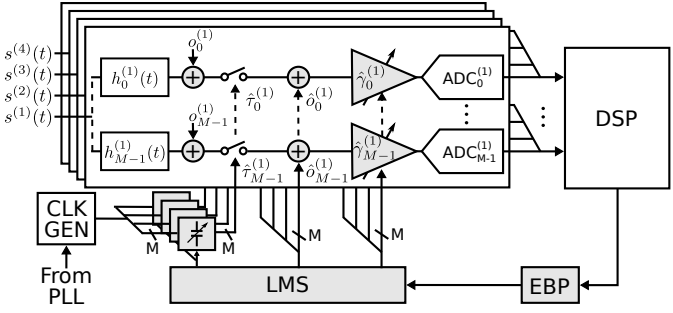}
  \caption{\label{f:ms_cal_sch}Block diagram of the mixed-signal
    calibration variant. The calibration with analog elements enables
    power consumption reduction.  The EBP is the same as the
    all-digital variant.}
\end{figure}

A mixed-signal based calibration technique can be also derived from
the error backpropagation (EBP) algorithm described in the previous
section.  Toward this end, sampling phase, gain, and offsets are
adjusted before the ADC\footnote{Compared
  to the full-digital compensation architecture, notice that the
  described mixed-signal solution is not able to compensate some
  effects such as bandwidth mismatches.} by using the gradient of the backpropagated
slicer error as depicted in Fig. \ref{f:ms_cal_sch}. Similarly to the full
digital solution, the DC offsets in the mixed-signal calibration
approach are compensated by using \eqref{eq:tildeo}. The gain is
iteratively adjusted by using
\begin{equation}
  \hat \gamma^{(i)}_{m,p+1}=\hat \gamma^{(i)}_{m,p}- \mu_{\gamma}
  {\hat e}^{(i)}[m+kM]w^{(i)}[m+kM],\quad \forall k,
\end{equation}
where $m=0,\cdots,M-1$ and $i=1,2,3,4$. Finally, since the
backpropagated slicer error is available at the ADC outputs, the
sampling phase can be iteratively adjusted by using the \emph{MMSE
  timing recovery algorithm}~\cite{lee_digital_2004}, i.e.,
\begin{align}
  \hat \tau^{(i)}_{m,p+1}=&\hat \tau^{(i)}_{m,p}-\mu_{\tau}{\hat e}^{(i)}[m+kM]\times\\
  \nonumber
                          &\left(w^{(i)}[m+kM+1]-w^{(i)}[m+kM-1]\right),\quad
                            \forall k
\end{align}
with $m=0,\cdots,M-1$. The calibration algorithm adjusts analog
elements already present in most implementations of the TI-ADC
\cite{reyes_design_2017,
  reyes_energy-efficient_2019,kull_24-72-gs/s_2018}. The clock
sampling phase is adjusted with variable delay lines, gain and offset
can be corrected in the comparator or with programmable gain
amplifiers (PGA), if needed.

\section{Simulation Results}
\label{s:sim_res}

\begin{table}[]
\centering
\caption{Parameters Used in Simulations (UDRVD: Uniformly Distributed
  Random Variable. VFS: Full-Scale Voltage).}
\label{t:sim_parameters}
  \begin{tabular}{l|c}
    \hline
    {\textbf{Parameter}}                                & \textbf{Value}
    \\ \hline 
    Modulation                                        & 16-QAM         \\ 
    Symbol Rate ($f_B=1/T$)                           & 96 GBd         \\ 
    Receiver Oversampling Factor ($T/T_s$)            & 4/3            \\ 
    Fiber Length                                      & 100 km \\ 
    Differential Group Delay (DGD)                    & 10 ps          \\ 
    Second Order Pol. Mode Disp. (SOPMD) & 1000 ps$^2$    \\ 
    Speed of Rotation of the Pol. at the Tx   & 2 kHz           \\ 
    Speed of Rotation of the Pol. at the Rx   & 20 kHz          \\ 
    TI-ADC Resolution                                 & 8 bit          \\ 
    TI-ADC Sampling Rate (all interleaves)            & 128 GS/s        \\ 
    Number of Interleaves of TI-ADC ($M$)             & 16             \\ 
    Number of Taps of CE ($L_g$)  & 7\\
    Rolloff Factor  & 0.10\\
    Nominal BW of Analog Paths ($B_0$) (see \eqref{eq:B}) & 53 GHz\\ 
    Gain Errors (see \eqref{eq:gain_error}) - UDRV & $\Delta_{\gamma^{(i)}_m}\in [\pm 0.15]$ \\
    Sampling Phase Errors - UDRV  &  $\delta_m^{(i)} \in [\pm 0.10]T$\\
    Bandwidth Mismatches (see \eqref{eq:B}) - UDRV &  $\Delta_{B_{m}^{(i)}} \in [\pm 0.075]B_0$\\
    I/Q Time Skew (see \eqref{eq:tau}) - UDRV &  $\tau_H,\tau_V\in [\pm 0.10]T$\\
    DC Offsets - UDRV  &  $o_m^{(i)} \in [\pm 0.025]$VFS\\
    \\ \hline
  \end{tabular}
\end{table}
\begin{figure}[t]
  \centering
  \includegraphics[width=\columnwidth]{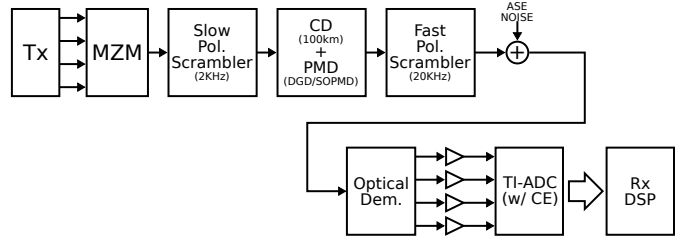}
  \caption{\label{f:simulation_setup}Block diagram of the system model
    used in the simulations.}
\end{figure}
The performance of the proposed backpropagation based adaptive CE is
investigated by running Montecarlo simulations of the setup shown in
Fig.~\ref{f:simulation_setup} and defined in Table
\ref{t:sim_parameters}.  Each test consists of 500 cases where the
impairment parameters are obtained by using uniformly distributed
random variables (UDRV). The electrical analog path responses
\eqref{eq:eq2} are simulated with first-order lowpass filters with
3dB-bandwidth defined by
\begin{equation}
  \label{eq:B}
  B^{(i)}_m=B_0+\Delta_{B^{(i)}_m},\quad i=1,2,3,4;\quad 
  m=0,\cdots,M-1, 
\end{equation}
where $B_0$ is the nominal BW and $\Delta_{B^{(i)}_m}$ is the BW
mismatch. Let $\tau_m^{(i)}$ be the mean group delay of the filter of
the $m$-th channel and $i$-th component. The impact of the I/Q time
skew of polarizations $H$ and $V$ defined as
\begin{equation}
  \label{eq:tau}
  \tau_H={\overline \tau}^{(1)}-{\overline \tau}^{(2)},\quad \tau_V={\overline \tau}^{(3)}-{\overline \tau}^{(4)}
\end{equation}
with ${\overline \tau}^{(i)}=\frac{1}{M}\sum_{m=0}^{M-1}\tau_m^{(i)}$,
is also investigated. We consider a 16-QAM modulation scheme with a
symbol rate of $1/T=\SI{96}{\giga\baud}$. Raised cosine filters with
rolloff factor $0.10$ for transmit pulse shaping are simulated (i.e.,
the nominal BW of the channel filters is
$B_0=1.1\times \frac{96}{2}\approx 53$ GHz). The optical
signal-to-noise ratio (OSNR) is set to that required to achieve a
bit-error-rate (BER) of $\sim 1.2\times 10^{-3}$ (see
\cite{freude_quality_2012,chan_optical_2010} for the definition of
OSNR). The oversampling factor in the DSP blocks is $T/T_s=4/3$. The
fiber length is \SI{100}{\km} with \SI{10}{\ps} of differential group
delay (DGD) and \SI{1000}{ps^2} of second-order PMD (SOPMD). Rotations
of the state of polarization (SOP) of \SI{2}{\kHz} and \SI{20}{\kHz}
are included at the transmitter and receiver, respectively.
TI-ADCs with 8-bit resolution, \SI{128}{\giga\sample\per\s} sampling
rate, and $M=16$ are simulated. The number of taps of the digital
compensation filters is $L_g=7$.

\subsection{Montecarlo Simulations of the Adaptive CE}
\label{sec:montecarlo}
\begin{figure}
\centering
\includegraphics[width=.87\columnwidth]{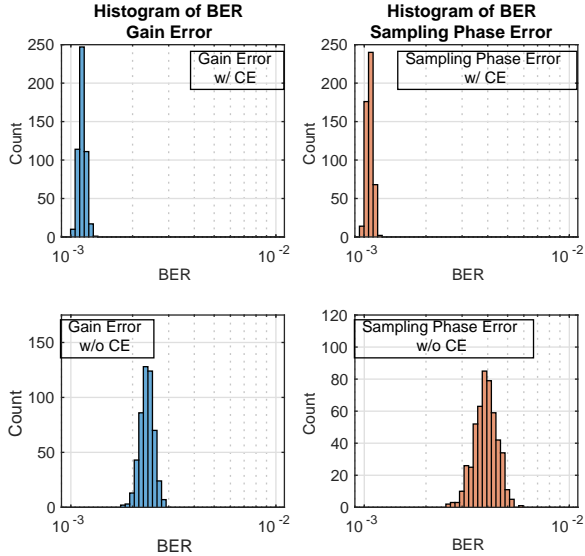}
\caption{\label{f:hist1}Histogram of the BER for 500 random cases with
  and without CE for a reference BER of $\sim 1.2\times
  10^{-3}$. Left: gain errors (only). Right: sampling phase errors
  (only). See simulation parameters in Table \ref{t:sim_parameters}.}
\end{figure}

\begin{figure}
\centering
\includegraphics[width=.87\columnwidth]{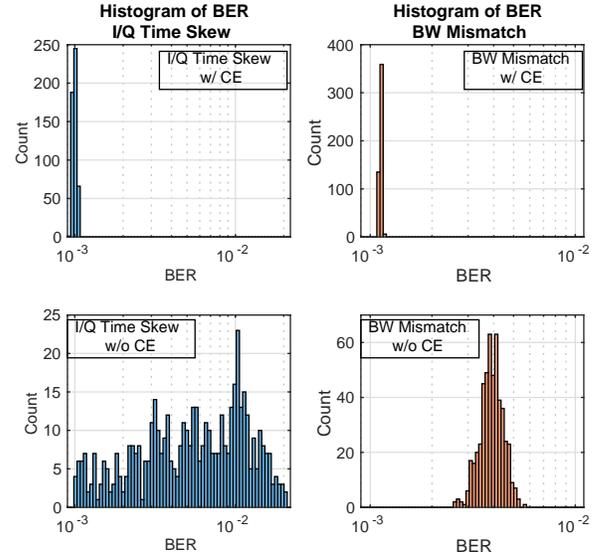}
\caption{\label{f:hist2}Histogram of the BER for 500 random cases with
  and without CE for a reference BER of $\sim 1.2\times
  10^{-3}$. Left: I/Q time skew (only). Right: BW mismatch (only).
  See simulation parameters in Table \ref{t:sim_parameters}.}
\end{figure}

Figs. \ref{f:hist1} and \ref{f:hist2} show the histograms of the BER
for the receiver with and without the CE in the presence of gain
errors, phase errors, I/Q time skew, and BW mismatches. Only one
effect is exercised in each case. Results of 500 random gain and phase
errors uniformly distributed in the interval
$\Delta_{\gamma^{(i)}_m} \in [\pm 0.15]$ (see \eqref{eq:gain_error})
and $\delta_m^{(i)}\in [\pm 0.10]T$, respectively, are depicted in
Fig. \ref{f:hist1}, whereas Fig. \ref{f:hist2} shows results for 500
random BW mismatches (see \eqref{eq:B}) and I/Q time skews (see
\eqref{eq:tau}) uniformly distributed in the interval
$\Delta_{B^{(i)}_m} \in [\pm 0.075]B_0$ and
$\tau_H,\tau_V\in [\pm 0.10]T$, respectively. In all cases, it is
observed that the proposed compensation technique is able to mitigate
the impact of all impairments when they are exercised
separately\footnote{Similar performance has been verified with random
  DC offsets \cite{solis_background_2020}.}. In particular, notice
that the proposed CE with $L_g=7$ taps practically eliminates the
serious impact on the receiver performance of the I/Q time skew values
of Table \ref{t:sim_parameters}.

\begin{figure}
\centering
\includegraphics[width=.9\columnwidth]{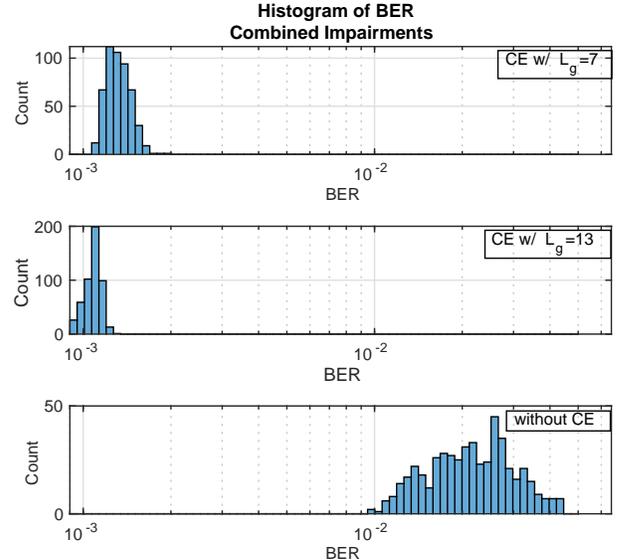}
\caption{\label{f:hist3}Histogram of the BER for 500 random cases with
  combined impairments as defined in Table
  \ref{t:sim_parameters}. Reference BER of $\sim 1.2\times
  10^{-3}$. Top: CE w/$L_g=7$ taps. Middle: CE w/$L_g=13$
  taps. Bottom: without CE.}
\end{figure}

Fig. \ref{f:hist3} shows histograms of the BER for the receiver with
and without the CE in the presence of the combined effects. Results of
500 cases with random gain errors, sampling phase errors, I/Q time
skews, BW mismatches, and DC offsets as defined in Table
\ref{t:sim_parameters}, are presented. Performance of the CE with
$L_g=13$ taps is also depicted. As before, note that the CE is able to
compensate the impact of all combined impairments. Moreover, note that
a slight performance improvement can be achieved when the number of
taps $L_g$ increases from 7 to 13.


\begin{figure}[t]
\centering
\includegraphics[width=1.\columnwidth]{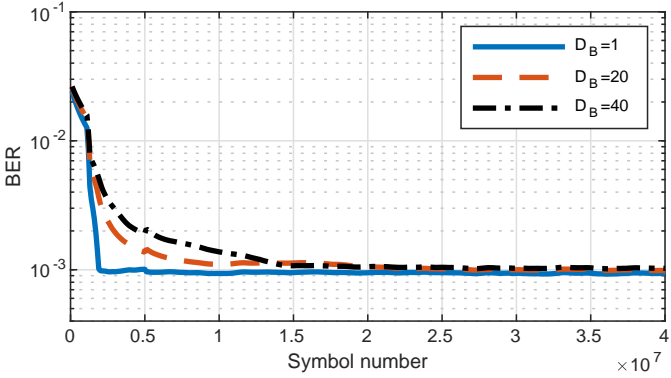}
\caption{\label{f:ber_conv} Convergence of the CE in the presence
  combined impairments for different block decimation factors $D_B$
  with $N=8192$.}
\end{figure}

As mentioned in Section \ref{s:bp_formulation}, the impairments of the
AFE and TI-ADCs change very slowly over time in multi-gigabit optical
coherent transceivers. Therefore the coefficient updates given by
\eqref{eq:lmsg2} and \eqref{eq:tildeo} do not need to operate at full
rate, and subsampling can be applied. Block processing and frequency
domain equalization based on the Fast Fourier Transform (FFT) are
widely used to implement high-speed coherent optical transceivers
\cite{morero_design_2016}. Then we propose to use block decimation of
the error samples to update the CE. Let $N$ be the block size in
samples to be used for implementing the EBP. Define $D_B$ the block
decimation factor. In this way, only one block of $N$ consecutive
samples of the oversampled slicer error \eqref{eq:oe} every $D_B$
blocks, i.e.,
\begin{equation}
  e^{(i)}[k ND_B+n],\quad n=0,1,\cdots,N-1,\forall k
\end{equation}
with $k$ integer, is used to adapt the CE. Fig. \ref{f:ber_conv}
depicts an example of the temporal evolution of the BER in the
presence of combined impairments according to Table
\ref{t:sim_parameters} for different values of the block decimation
factor $D_B$ with $N=8192$. The instantaneous BER is evaluated every
$10^5$ symbols and then processed by a moving average filter of size
$40$. Gear shifting is used to accelerate the convergence of the CE
and reduce the steady-state MSE. In all cases, notice that the use of
block decimation practically does not impact on the resulting
BER. Therefore it can be adopted to drastically reduce the
implementation complexity, as shall be discussed in Section
\ref{s:complexity}.

\subsection{Mixed-Signal Compensation of TI-ADC with Highly
  Interleaved Architectures}
\label{ss:high_interleaved_sim}

\begin{figure}
\centering
\includegraphics[width=1.\columnwidth]{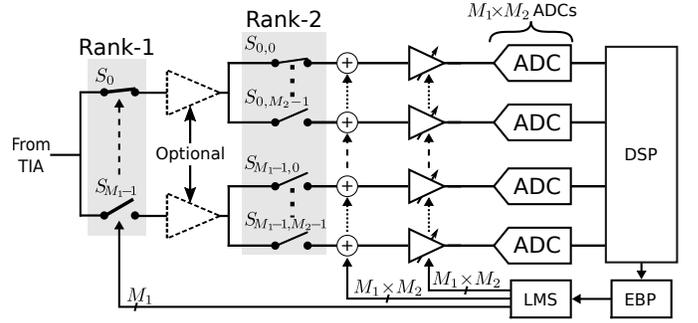}
\caption{\label{f:sch_hier_tiadc} Example of application of the
  proposed backpropagation based mixed-signal calibration in a typical
  two-rank hierarchical TI-ADC.}
\end{figure}

The performance of the mixed-signal scheme of Section \ref{sec:mixed}
is investigated in typical hierarchical ultra high-speed TI-ADCs such
as those used in high speed
receivers~\cite{reyes_energy-efficient_2019, kull_24-72-gs/s_2018,
  kim_161-mw_2020}. This hierarchical TI-ADC architecture organizes
the T\&H in two or more ranks with a high number of
sub-ADCs. Fig.~\ref{f:sch_hier_tiadc} depicts an example with two
ranks. Rank 1 includes $M_1$ switches each of which feeds $M_2$ T\&H
stages of Rank 2. Then, $M_1\times M_2$ ADCs are used to digitize the
input signal. Successive approximation register (SAR) ADCs are used
for this application due to their power efficiency at the required
sampling rate and resolution. This approach relaxes the requirements
for the clock generation and synchronization. Furthermore, the impact
on the input bandwidth is reduced in contrast to T\&H with direct
sampling~\cite{greshishchev_40_2010}. As an example of application of
the mixed-signal compensation scheme of Section \ref{sec:mixed}, its
performance in a hierarchical TI-ADC with $M_1=16$ and $M_2=8$ (i.e.,
$M_1\times M_2=128$ individual converters) is evaluated. A clock
jitter of \SI{100}{\fs} RMS is added to this simulation. Notice that
the mixed-signal calibration algorithm adjusts the $M_1$ sampling
phases of the switches in the first rank, and the $M_1\times M_2$
gains and offsets of the individual sub-ADCs.

\begin{figure}
  \centering
  \includegraphics[width=.95\columnwidth]{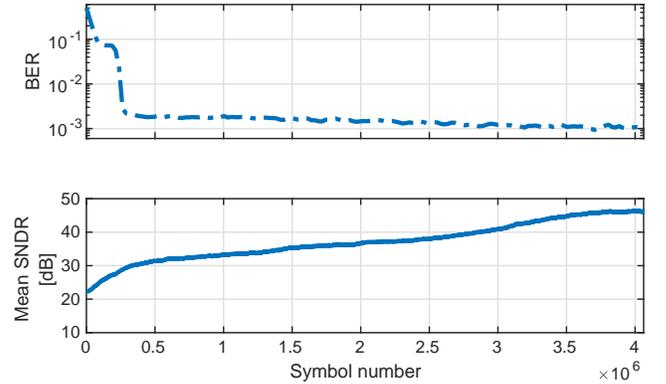}
\caption{\label{f:ber_evo_jitter} BER and SNDR evolution in a
  hierarchical TI-ADC based DP optical coherent receiver with the
  backpropagation based mixed-signal compensation in the presence
  combined impairments. $M_1=16$ and $M_2=8$.}
\end{figure}

Fig.~\ref{f:ber_evo_jitter} shows the temporal evolution of both the
BER and the mean \emph{signal-to-noise-and-distortion-ratio} (SNDR)
~\cite{reyes_energy-efficient_2019}. A slower convergence than the
previous simulation is observed as a result of the larger number of
converters (i.e., 128 vs 16). Nevertheless, we verify that the
proposed backpropagation based mixed-signal compensation scheme is
able to mitigate the impact of the impairments in hierarchical
TI-ADC. In particular, note that the SNDR can be improved from
$\sim 20$dB to $\sim 45$dB by using the proposed background
calibration technique.



\section{Hardware Complexity Analysis}
\label{s:complexity}
This section discusses some practical aspects of the implementation of
the proposed compensation technique. We focus on the two main blocks
of the all digital architecture: the compensation equalizer and the
error backpropagation block.

\subsection{Implementation of the Compensation Equalizer}

\begin{figure}[t]
  \centering
  \includegraphics[width=1.\columnwidth]{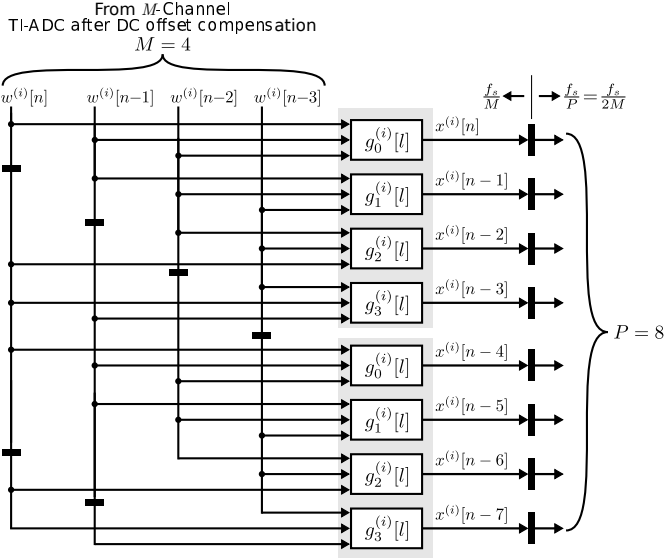}
  \caption{\label{f:g_par} Example of a parallel implementation of the
    CE with $M=4$, $L_g=3$, and parallelism factor $P=2M=8$. }
\end{figure}

As described in Section~\ref{sec:EBP}, the compensation equalizer in a
DP optical coherent receiver comprises 4 real valued finite impulse
response (FIR) filters $\tilde{g}_n^{(i)}[l]$ with $i=1,2,3,4$, and
$l=0,\cdots, L_g-1$. From computer simulations of
Section~\ref{s:sim_res} it was observed that $L_g=7$ is enough to
properly compensate the AFE and TI-ADC impairments. Therefore a time
domain implementation is preferred for the CE.  Each of these filters
has $M$ independent impulse responses $g_n^{(i)}[l]$ which are time
multiplexed as $\tilde{g}_n^{(i)}[l]=g_{\lfloor n\rfloor_M}^{(i)}[l]$
(see \eqref{eq:gPC}). Note that time multiplexing of filters with
independent responses does not translate to additional complexity when
the filter is implemented with a parallel architecture. The use of
parallel implementation is mandatory in high speed optical
communication where parallelism factors on the order of 128 o higher
are typical. In these architectures, the parallelism factor $P$ can be
chosen to be a multiple of the ADC parallelism factor $M$, i.e.,
$P=q\times M$ where $q$ is an integer. Therefore the different time
multiplexed coefficients are used in fixed positions of the
parallelism without incurring in significant additional complexity in
relation to a filter with just one set of coefficients (see
Fig. \ref{f:g_par}). We highlight that the resulting filter is
equivalent in complexity to the I/Q-skew compensation filter already
present in current coherent receivers~\cite{morero_design_2016}. Since
the proposed scheme also corrects skew, the classical skew correction
filter can be replaced by the proposed CE without incurring
significant additional area or power.

\subsection{Implementation of the Error Backpropagation Block}

A straightforward implementation of error backpropagation must include
a processing stage for each DSP block located between the ADCs and the
slicers. Typically these blocks comprise the BCD, FFE, TR
interpolators, and the FCR. All these blocks can be mathematically
modeled as a sub-case of the generic receiver DSP block used in
Section~\ref{s:bp_formulation} and the Appendix. The EBP block is
algorithmically equivalent to its corresponding DSP block with the
only difference that the coefficients are \emph{transposed} (i.e.,
compare \eqref{eq:un} and \eqref{eq:epb}). Therefore, in the worst
case, the EBP complexity would be similar to that of the receiver DSP
block\footnote{Note that the LMS adaptation hardware of the FFE, the
  PLL of the FCR and the PLL of the TR do not need to be implemented
  in the EBP path, which further reduces the complexity of the
  latter.}. Since doubling power and area consumption is not
acceptable for commercial applications, important simplifications must
be provided.

Considering that AFE and TI-ADC impairments change very slowly over
time in multi-gigabit optical coherent transceivers, the coefficient
updates given by \eqref{eq:lmsg2} and \eqref{eq:tildeo} do not need to
operate at full rate, and subsampling can be applied. The latter
allows implementation complexity, and particularly power dissipation,
to be drastically reduced. In Section \ref{sec:montecarlo} we
evaluated the performance with block decimation where one block of $N$
consecutive samples of the oversampled slicer error are used every
$D_B$ blocks. Simulation results not included here have shown a good
performance even with $N=8192$ and $D_B=256$. The block based
decimation approach allows the EBP algorithm to be implemented in the
frequency domain when necessary to reduce complexity (for example in
the EBP of the BCD and FFE). This error decimation reduces the power
dissipation of the EBP to only $1/D_B$ of the power of the
corresponding DSP blocks, equivalent to less than 1\% in the simulated
example. However, the areas of the EBP blocks are still equivalent to
the area of their corresponding DSP blocks. To reduce area, the EBP
blocks could be implemented using a serial
architecture\footnote{Typically, a serial implementation requires that
  hardware such as multipliers be reused with variable numerical
  values of coefficients, whereas in a parallel implementation
  hardware can be optimized for fixed coefficient values. This results
  in a somewhat higher power per operation in a serial
  implementation. Nevertheless, the drastic power reduction achieved
  through decimation greatly outweighs this effect.} or a lower
parallelism factor. If a serial implementation is chosen, an area
reduction proportional to the parallelism factor is expected at the
expense of increasing the latency by a similar amount. The resulting
latency is $2 \times (N_{BCD}+N_{FFE}) \times P$ samples, where
$N_{BCD}$ and $N_{FFE}$ are the block sizes of the FFTs used to
implement the BCD and FFE, respectively (factor 2 includes the FFT /
IFFT pair). The latencies of the EBP blocks for the TR interpolators
and FCR can be neglected. Therefore the CE adaptation speed is not
reduced by a serial implementation of the EBP blocks if
$2 \times (N_{BCD}+N_{FFE}) \times P < N \times D_B$. Details of
efficient architectures for implementing the error backpropagation
block will be addressed in a future work.

\section{Conclusions}
\label{s:conclusion}
A new TI-ADC background calibration algorithm based on the
backpropagation technique has been presented in this paper.  Two
implementation variants were presented, one of them all-digital and
the other mixed-signal. Simulation results have shown a fast, robust
and almost ideal compensation/calibration of TI-ADC sampling time,
gain, offset, and bandwidth mismatches as well as I/Q time skew
effects under different test conditions in the example of application
of a DSP-based optical coherent receiver. Hardware complexity is
minimized with serial processing and decimation. As the technique runs
in background, the calibration can track parameter variations caused
by temperature, voltage, aging, etc., without operational
interruptions.

\section*{Acknowledgements}
The authors would like to thank Dr. Ariel Pola for his helpful advice
on various technical issues related to the hardware implementation.

\section*{Appendix}
In this Appendix the stochastic gradient of the squared error defined
by \eqref{eq:grad} is derived. The total squared error \eqref{eq:eT}
is
\begin{equation}
  \label{eq:eT2}
  {\mathcal E}_k=\sum_{j=1}^4 \left|e_k^{(j)}\right|^2=\sum_{j=1}^4
  \left(u_k^{(j)}- \hat a_k^{(j)}\right)^2.
\end{equation}
with $u_k^{(j)}$ given by \eqref{eq:u1}. Define the \emph{average}
squared error as
\begin{equation}
  \label{eq:mse}
  \overline {\mathcal E}_N=\frac{1}{2N+1}\sum_{k=-N}^{N}\sum_{j=1}^4
  \left(u_k^{(j)}-\hat a_k^{(j)}\right)^2.
\end{equation}

The derivative of $\overline {\mathcal E}_N$ with respect to
$g^{(i_0)}_{m_0}[l_0]$ is
\begin{equation}
  \label{eq:dEdg}
  \frac{\partial { \overline {\mathcal E}_N}}{\partial g^{(i_0)}_{m_0}[l_0]}=\frac{2}{2N+1}\sum_{k=-N}^{N}\sum_{j=1}^4
  e_k^{(j)}\frac{\partial u_k^{(j)}}{\partial g^{(i_0)}_{m_0}[l_0]},
\end{equation}
where $l_0\in\{0, 1, \cdots,L_g-1\}$, $m_0\in\{0, 1, \cdots,M-1\}$,
and $i_0\in\{1,2,3,4\}$. From the slicer error $e_{k}^{(j)}$ given by
\eqref{eq:ePC}, define the $T_s=T/2$ oversampled slicer error as
\begin{equation}
  e^{(j)}[n] = 
  \begin{cases} 
    e_{n/2}^{(j)}              & \mbox{if } n= 0,\pm 2,\pm 4,\cdots   \\
    0 & \mbox{otherwise}
  \end{cases}.
\end{equation}
Thus, \eqref{eq:dEdg} can be rewritten as
\begin{equation}
  \label{eq:dEdg2}
  \frac{\partial { \overline {\mathcal E}_N}}{\partial g^{(i_0)}_{m_0}[l_0]}=\frac{2}{2N+1}\sum_{n=-2N}^{2N}\sum_{j=1}^4
  e^{(j)}[n]\frac{\partial u^{(j)}[n]}{\partial g^{(i_0)}_{m_0}[l_0]},
\end{equation}
where $u^{(j)}[n]$ is the oversampled compensation equalizer output
given by
\begin{equation}
  \label{eq:un}
  u^{(j)}[n]=\sum_{i=1}^4\sum_{l=0}^{L_{\Gamma}-1}
  {\Gamma}^{(j,i)}_{n}[l]{x}^{(i)}[n-l],\quad j=1,\cdots,4.
\end{equation}
The time index $n$ can be expressed as
\begin{equation}
  \label{eq:n}
  n=m+k'M,\quad m=0,1,\cdots,M-1;\quad \forall k',
\end{equation}
with $k'$ integer. Then, omitting the constant factor
$\frac{2}{2N+1}$, the derivative \eqref{eq:dEdg2} can be expressed as
\begin{equation}
  \label{eq:dEdg3}
  \frac{\partial { \overline {\mathcal E}_N}}{\partial g^{(i_0)}_{m_0}[l_0]}\propto\sum_{k'}\sum_{m=0}^{M-1}\sum_{j=1}^4
  e^{(j)}[m+k'M]\frac{\partial u^{(j)}[m+k'M]}{\partial g^{(i_0)}_{m_0}[l_0]}.
\end{equation}

Next we evaluate the derivative
$\frac{\partial u^{(j)}[m+k'M]}{\partial
  g^{(i_0)}_{m_0}[l_0]}$. Assuming that the DSP filter coefficients
$\Gamma^{(j,i)}_{n}[l]$ do not depend on $g^{(i_0)}_{m_0}[l_0]$, from
\eqref{eq:un} and \eqref{eq:n} we verify that
\begin{equation}
  \label{eq:dundg}
  \frac{\partial u^{(j)}[m+k'M]}{\partial g^{(i_0)}_{m_0}[l_0]}=\sum_{i=1}^4\sum_{l=0}^{L_{\Gamma}-1}
  {\Gamma}^{(j,i)}_{m+k'M}[l]\frac{\partial {x}^{(i)}[m+k'M-l]}{\partial g^{(i_0)}_{m_0}[l_0]}.
\end{equation}

Based on \eqref{eq:n}, the signal at the DSP block input $i$ given by
\eqref{eq:eq6b} can be rewritten as
\begin{equation}
  x^{(i)}[m+k'M]=\sum_{l'=0}^{L_g-1} {g}^{(i)}_{m}[l']
  {w}^{(i)}[m+k'M-l'].
\end{equation}
Therefore,
\begin{equation}
  \label{eq:dxdg}
  \frac{\partial x^{(i)}[m+k'M]}{\partial g^{(i_0)}_{m_0}[l_0]}=
  {w}^{(i)}[m+k'M-l_0]\delta_{m,m_0}\delta_{i,i_0},
\end{equation}
where $\delta_{n,m}$ is the Kronecker delta function (i.e.,
$\delta_{n,m}=1$ if $n=m$ and $\delta_{n,m}=0$ if $n\ne m$). Replacing
\eqref{eq:dxdg} in \eqref{eq:dundg} we get
\begin{align}
  \nonumber
  &\frac{\partial u^{(j)}[m+k'M]}{\partial g^{(i_0)}_{m_0}[l_0]}=\\
  \label{eq:dudg2}
  &\quad\quad\quad \sum_{l=0}^{L_{\Gamma}-1} {\Gamma}^{(j,i_0)}_{m+k'M}[l] {w}^{(i_0)}[m+k'M-l-l_0]\delta_{m,m_0}.
\end{align}
Using \eqref{eq:dudg2} in \eqref{eq:dEdg3}, we obtain
\begin{align}
  \frac{\partial { \overline {\mathcal E}_N}}{\partial g^{(i_0)}_{m_0}[l_0]}\propto&\sum_{k'}\sum_{j=1}^4
  e^{(j)}[m_0+k'M]\times\\
\nonumber
    &\sum_{l=0}^{L_{\Gamma}-1} {\Gamma}^{(j,i_0)}_{m_0+k'M}[l] {w}^{(i_0)}[m_0+k'M-l-l_0].
\end{align}
Finally, we set $kM=k'M-l$ resulting
\begin{equation}
  \label{eq:dEdg4}
  \frac{\partial { \overline {\mathcal E}_N}}{\partial g^{(i_0)}_{m_0}[l_0]}\propto\sum_{k}{\hat e}^{(i_0)}[m_0+kM] {w}^{(i_0)}[m_0+kM-l_0],
\end{equation}
where
\begin{equation}
  \label{eq:epb}
  {\hat e}^{(i)}[n]=\sum_{j=1}^4\sum_{l=0}^{L_{\Gamma}-1} {\Gamma}^{(j,i)}_{n+l}[l]e^{(j)}[n+l]
\end{equation}
is the backpropagated error. Notice that \eqref{eq:dEdg4} is the
average of the instantaneous gradient component given by
${\hat e}^{(i_0)}[m_0+kM] {w}^{(i_0)}[m_0+kM-l_0]$. Therefore, an
instantaneous gradient of the square error can be obtained as
\begin{equation}
  \nabla_{{\bold g}^{(i)}_{m}} {\mathcal E}_k\propto{\hat e}^{(i)}[m+kM]{\bold w}^{(i)}[m+kM],
\end{equation}
where ${\bold w}[n]$ is the $L_g$-dimensional vector with the samples
at the CE input defined by \eqref{eq:vecw}.

\bibliographystyle{IEEEtran/IEEEtran}
\balance
\bibliography{IEEEtran/IEEEabrv,TIADC_BP_CAL_TCAS2020}

\end{document}